\documentclass[fp,twocolumn]{jpsj3}
\usepackage{txfonts}

\title{Phase transition of vortex states in two-dimensional superconductors under a oscillating magnetic field from the chiral helimagnet}

\author{Saoto Fukui$^1$, Masaru Kato$^1$, Yoshihiko Togawa$^2$, Osamu Sato$^3$}
\inst{${}^1$Department of Mathematical Sciences, Osaka Prefecture University, 1-1, Gakuencho, Nakaku, Sakai, Osaka 599-8531, Japan \\
${}^2$Department of Physics and Electronics, Osaka Prefecture University, 1-1, Gakuencho, Nakaku, Sakai, Osaka 599-8531, Japan \\
${}^3$Osaka Prefecture University College of Technology, 26-12, Saiwaicho, Neyagawa, Osaka 572-8572, Japan} 

\abst{We have investigated vortex states in two-dimensional superconductors under a oscillating magnetic field from a chiral helimagnet.
We have solved the two-dimensional Ginzburg-Landau equations with finite element method.
We have found that when the magnetic field from the chiral helimagnet increases, vortices appear all at once in all periodic regions.
This transition is different from that under the uniform magnetic field.
Under the composite magnetic field with the oscillating and uniform fields (down-vortices), vortices antiparallel to the uniform magnetic field disappear.
Then, the small uniform magnetic field easily remove down-vortices.
}

\begin{document}
\maketitle

\section{Introduction}
Vortex states in a type-II superconductor are affected by many factors such as magnetic field, current and so on.
When a homogeneous magnetic field is applied to a superconductor, quantum fluxes appear and form the Abrikosov lattice \cite{Abrikosov}, but the vortex configuration is different from the Abrikosov lattice under an inhomogeneous magnetic field.
This inhomogeneous magnetic field can be created by a ferromagnet.

In a ferromagnet / superconductor bilayer structure, vortices appear spontaneously because of interaction between magnetic fluxes of vortices in the superconductor and magnetic domains in the ferromagnet \cite{FM_SC_Hybrid, FM_SC_FSB, FM_SC_FSB_ex}.
Moreover, it is known that this hybrid structure enhances a critical current in the superconductor because of a pinning effect of magnetic domain structures on vortices.
Because of these strong correlations between the magnetic structure and the vortex structure in the superconductor, if the configuration of magnetic moments in the magnet changes, the vortex structure in the superconductor may be affected by the magnet.
As the spatially varied magnetic materials, we take a chiral helimagnet.

\begin{figure}[htbp]
 \includegraphics[scale=0.38]{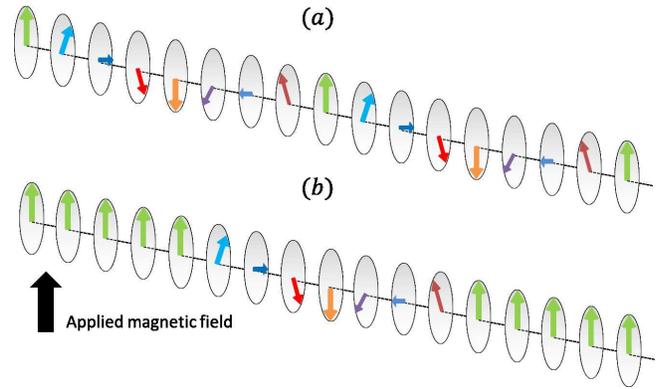}
 \caption{(a) The magnetic structure in the chiral helimagnet, (b) the magnetic structure in chiral soliton lattice.}
 \label{CHM_CSL}
\end{figure}

In the chiral helimagnet, there are two kinds of interactions between nearest neighbor spins, the ferromagnetic exchange interaction and the antisymmetric Dzyaloshinsky-Moriya (DM) interaction \cite{Dzyaloshinsky, Moriya}.
The ferromagnetic interaction causes directions of nearest neighbor spins to be parallel, while the DM interaction causes directions of nearest neighbor spins to be perpendicular.
The DM interaction breaks the rotational symmetry, then the chiral helimagnet has chirality.
The direction of the rotation is determined by a crystal structure.
In the chiral helimagnet, the competition between the ferromagnetic exchange interaction and the antisymmetric Dzyaloshinsky-Moriya (DM) interaction make a helical rotation of magnetic moments, which is shown in Fig. \ref{CHM_CSL}(a).

In the chiral helimagnet, when the magnetic field is applied perpendicular to the helical axis, the helical configuration of magnetic moments changes into a periodic soliton structure, which is shown in Fig. \ref{CHM_CSL}(b).
This periodic soliton structure is called chiral soliton lattice.
The transition from the helical magnetic structure to chiral soliton lattice is observed in CrNb$_3$S$_6$ experimentally \cite{CHM_ex_1, CHM_ex_2}.
It is shown theoretically that the magnetic structure in the chiral helimagnet has characteristic properties \cite{CHM_theor_1}.
In particular, the application of chiral helimagnets to the spintronics is discussed theoretically \cite{CHM_theor_2}.

We expect that peculiar magnetic structures in the chiral helimagnet affect the superconductor strongly.
These effects of the chiral helimagnet may be very different from those of the ferromagnet.
So, we investigate effects of the chiral helimagnet on the superconductor.
In particular, we focus on effects on vortices in the superconductor.

In this paper, we show effects of the chiral helimagnet on vortices in the superconductor theoretically.
In section 2, we introduce our model, the Ginzburg-Landau equations for our model in the superconductor, and a numerical method.
In section 3, we show vortex states under the magnetic field from the chiral helimagnet.
In section 4, we give a brief summary of this paper.

\section{Method}
We consider a two-dimensional superconducting film under a magnetic field from the chiral helimagnet.
In the two-dimensional superconductor model, only a magnetic field normal to the superconducting film is effective.
So, we consider only an oscillating magnetic field normal to the superconducting film.
Under this magnetic field, the order parameter of the superconductor is expected to vary spatially \cite{Fukui_sust}.
In order to incorporate the spatial variation of the order parameter, we solve the Ginzburg-Landau equations.

We start the Ginzburg-Landau free energy.
\begin{eqnarray}
 & & \mathcal{F}(\psi,\mbox{\boldmath $A$}) =  \int_{\Omega} \left( f_n + \alpha |\psi|^2 + \frac{\beta}{2} |\psi|^4 \right) d\Omega \nonumber \\
 & & + \int_{\Omega} \left\{ \frac{1}{2m^\ast} \left| \left( -i\hbar \nabla - \frac{e^\ast \mbox{\boldmath $A$}}{c} \right) \psi \right|^2 + \frac{|\hbar|^2}{8\pi} - \frac{\mbox{\boldmath $h$} \cdot \mbox{\boldmath $H$}_{\rm ext}}{4\pi} \right\}, \nonumber \\  \label{gl_free_start}
\end{eqnarray}
where $\psi$ is a superconducting order parameter, 
$m^\ast$ is an effective mass of the superconductor, and $e^\ast$ is an effective charge of electrons in the superconductor.
$\alpha(T) = \alpha' (T-T_c)$ is a coefficient where $\alpha'>0$ is a constant and $T$ is a temperature and $T_c$ is a critical temperature.
$\beta$ is a positive constant and $\mbox{\boldmath $h$} = \nabla \times \mbox{\boldmath $A$}$ is a microscopic magnetic field where $\mbox{\boldmath $A$}$ is a magnetic vector potential.
$\mbox{\boldmath $H$}_{\rm ext}$ is an external magnetic field, which is given by,
\begin{equation}
 \mbox{\boldmath $H$}_{\rm ext} = \mbox{\boldmath $H$}_{\rm CHM} + \mbox{\boldmath $H$}_{\rm appl}, \label{field_ext}
\end{equation}
where $\mbox{\boldmath $H$}_{\rm CHM}$ is a magnetic field from the chiral helimagnet and $\mbox{\boldmath $H$}_{\rm appl}$ is the uniform applied magnetic field.
We perform Legendre transformation to the free energy, which leads to,
\begin{eqnarray}
 \mathcal{G}(\psi,\mbox{\boldmath $A$}) &=& \int d\Omega \left[ \frac{1}{2} \left( \sqrt{\beta} |\psi|^2 + \frac{\alpha}{\sqrt{\beta}} \right)^2 \right. \nonumber \\
                                        & & \left. + \frac{1}{2m^\ast} \left| \left( -i\hbar\nabla + \frac{e^\ast \mbox{\boldmath $A$}}{c} \right) \psi \right|^2 + \frac{|\mbox{\boldmath $h$} - \mbox{\boldmath $H$}_{\rm ext}|^2}{8\pi} \right. \nonumber \\
                                        & & \left. + \frac{1}{8\pi} ({\rm div}~\mbox{\boldmath $A$})^2 \right] \label{gl_free}.
\end{eqnarray}
We add the last term $({\rm div}~\mbox{\boldmath $A$})^2/(8\pi)$ in order to insure the London Gauge, ${\rm div}~\mbox{\boldmath $A$}=0$.
To minimize the GL free energy, we use the Frech$\acute{\rm e}$t derivative;
\begin{eqnarray}
 \frac{\mathcal{G}(\psi+\epsilon\tilde{\psi},\mbox{\boldmath $A$}) - \mathcal{G}(\psi,\mbox{\boldmath $A$})}{\epsilon}                    &=& 0 \label{derivative_psi}, \\
 \frac{\mathcal{G}(\psi,\mbox{\boldmath $A$} + \epsilon \tilde{\mbox{\boldmath $A$}}) - \mathcal{G}(\psi,\mbox{\boldmath $A$})}{\epsilon} &=& 0, \label{derivative_a}
\end{eqnarray}
where $\epsilon$ is a small parameter.
$\tilde{\psi}$ and $\tilde{\mbox{\boldmath $A$}}$ are variations, or the test functions of the order parameter and vector potential, respectively.
Then, we obtain following Ginzburg-Landau equations,
\begin{eqnarray}
 & & \int d\Omega \left[ (i\nabla \psi - \mbox{\boldmath $A$} \psi)(-i\nabla \tilde{\psi}^\ast - \mbox{\boldmath $A$} \tilde{\psi}^\ast) \right. \nonumber \\
 & & +(i\nabla \tilde{\psi} - \mbox{\boldmath $A$} \tilde{\psi})(-i\nabla \psi^\ast - \mbox{\boldmath $A$} \psi^\ast) \nonumber \\
 & & \left. + \frac{1}{\xi(T)^2} (|\psi|^2 - 1)(\psi \tilde{\psi}^\ast + \tilde{\psi} \psi^\ast) \right] = 0 \label{gl_1}, \\
 & & \int d\Omega \left[ \kappa^2 \xi(T)^2 \{ {\rm div}~\mbox{\boldmath $A$}~\cdot \tilde{\mbox{\boldmath $A$}} + (\nabla \times \mbox{\boldmath $A$})(\nabla \times \tilde{\mbox{\boldmath $A$}}) \} \right. \nonumber \\
 & & \left. + |\psi|^2 \mbox{\boldmath $A$} \cdot \tilde{\mbox{\boldmath $A$}} - \frac{i}{2} (\psi^\ast \nabla \psi - \psi \nabla \psi^\ast) \tilde{\mbox{\boldmath $A$}} \right] \nonumber \\
 & & = \kappa^2 \xi(T)^2 \int d\Omega \frac{2\pi}{\Phi_0} \mbox{\boldmath $H$}_{\rm ext} \cdot (\nabla \times \tilde{\mbox{\boldmath $A$}}) \label{gl_2},
\end{eqnarray}
where $\kappa = \lambda(T)/\xi(T),~\lambda(T),$ and $\xi(T)$ are the Ginzburg-Landau parameter, the penetration depth, and the coherence length, respectively.
$\Phi_0 = hc/2e$ is the quantum flux and $e$ is the electronic charge. 

In order to calculate the Ginzburg-Landau equations, we use the finite element method (FEM) \cite{Kato_FEM}.
In the two-dimensional FEM, we divide the superconducting region into triangular finite elements.
The order parameter $\psi$ and the vector potential $\mbox{\boldmath $A$}$ for $e$-th element are expanded with area coordinates,
\begin{eqnarray}
 \psi(x,y) &=& \sum_e \left\{ N_1^e(x,y) \psi_1^e + N_2^e(x,y) \psi_2^e + N_3^e(x,y) \psi_3^e \right\} \nonumber \\ \label{psi_fem} \\
 \mbox{\boldmath $A$}(x,y) &=& \sum_e \left\{ N_1^e(x,y) \mbox{\boldmath $A$}_1^e + N_2^e(x,y) \mbox{\boldmath $A$}_2^e + N_3^e(x,y) \mbox{\boldmath $A$}_3^e \right\}, \nonumber \\  \label{vec_fem} 
\end{eqnarray}
where $N_i^e,~\psi_i^e,~{\rm and}~\mbox{\boldmath $A$}_i^e~(i=1,~2,~{\rm and}~3)$ is the area coordinate, the order parameter, and the vector potential at nodes in $e-$th element, respectively. 
We substitute Eqs. (\ref{psi_fem}) and (\ref{vec_fem}) into the Ginzburg-Landau equations [Eqs. (\ref{gl_1}) and (\ref{gl_2})].
Then, we set test functions as,
\begin{eqnarray}
 \tilde{\psi} &=& N_i(x,y)~~(i=1,~2,~3), \label{test_psi} \\
 \tilde{\mbox{\boldmath $A$}} &=& N_i(x,y) \mbox{\boldmath $e$}_j~~(i=1,~2,~3;~~j=x,y) \label{test_vec}. 
\end{eqnarray}
We obtain the Ginzburg-Landau equations in the FEM as,
\begin{eqnarray}
 & & \sum_j \left[ P_{ij}^e( \{ \psi \}, \{ \mbox{\boldmath $A$} \}) + P_{ij}^{e2R}(\{ \psi \}) \right] {\rm Re}~\psi_j^e \nonumber \\
 & & + \sum_j \left[ Q_{ij}^2( \{ \mbox{\boldmath $A$} \}) + Q_{ij}^{e2}(\{ \psi \}) \right] {\rm Im}~\psi_j^e = V_i^{eR} (\{ \psi \}), \label{gl_fem_1} \\
 & & \sum_j \left[ -Q_{ij}^e(\{ \mbox{\boldmath $A$} \}) + Q_{ij}^{e2}(\{ \psi \}) \right] {\rm Re}~\psi_j^e \nonumber \\
 & & + \sum_j \left[ P_{ij}^e(\{ \mbox{\boldmath $A$} \}) + P_{ij}^{2I}(\{ \psi \}) \right] {\rm Im}~\psi_j^e = V_i^{eI} (\{ \psi \}), \label{gl_fem_2} \\
 & & \sum_j R_{ij}^e(\{ \psi \}) A_{jx}^e + \sum_j S_{ij}^e A_{jy}^e = T_i^{ex}(\{\psi\}) - U_i^{ey}, \label{gl_fem_3} \\
 & & -\sum_j S_{ij}^e A_{jx}^e + \sum_j R_{ij}^e (\{\psi\}) A_{jy}^e = T_i^{ey} (\{\psi\}) + U_i^{ex}. \nonumber \\ \label{gl_fem_4} 
\end{eqnarray}
We define coefficients in the reference \cite{Fukui_sust}.

Next, we assume that distribution of the magnetic field $\mbox{\boldmath $H$}_{\rm CHM}$ is proportional to the distribution of magnetic moments in the chiral helimagnet.
As discussed above, in the two-dimensional superconductor, only the $z-$component magnetic field is effective.
Then, the magnetic field $(\mbox{\boldmath $H$}_{\rm CHM})_z$ can be expressed by the analytical solution from the Hamiltonian of the chiral helimagnet \cite{Kishine_CHM}.

\begin{figure}[t]
 \includegraphics[scale=0.3]{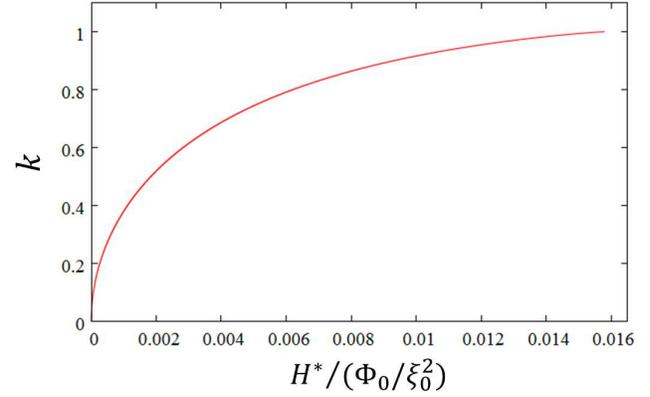}
 \caption{Dependence of modulus of the elliptic function $k$ on the applied magnetic field for $|\mbox{\boldmath $D$}|/J=0.16$.}
 \label{k_appl}
\end{figure}

The Hamiltonian of the chiral helimagnet is given as,
\begin{equation}
 \mathcal{H} = -J \sum_i \mbox{\boldmath $S$}_i \cdot \mbox{\boldmath $S$}_{i+1} + \mbox{\boldmath $D$} \cdot \sum_i \mbox{\boldmath $S$} \times \mbox{\boldmath $S$}_{i+1} - 2\mu_B \mbox{\boldmath $H$}_{\rm appl} \cdot \sum_i \mbox{\boldmath $S$}_i \label{hamiltonian_chm}
\end{equation}
Here, $S_i$ is the $i-$th spin. The first term is the ferromagnetic exchange interaction with strength $J$.
$\mbox{\boldmath $D$}$ is Dzyaloshinsky-Moriya vector of the Dzyaloshinsky-Moriya interaction.
The second term is the Dyzloshinsky-Moriya interaction with the Dzyaloshinsky-Moriya vector $\mbox{\boldmath $D$}$.
The last term is the Zeeman interaction and $\mu_B$ is the Bohr magneton.
The $i-$th spin $\mbox{\boldmath $S$}_i$ is expressed by polar coordinates as,
\begin{equation}
 \mbox{\boldmath $S$}_i = S(\sin{\theta}_i\cos{\theta}_0,~\sin{\theta}_i\sin{\theta}_0,~\cos{\theta}_i). \label{spin_polar}
\end{equation}
In the chiral helimagnet, $\theta_0 = \pi/2$.
Then, in the continuum limit, the energy of the chiral helimagnet becomes as,
\begin{equation}
 \mathcal{H} = \mathcal{J} S^2 \int dx \left[ \frac{1}{2} \left( \frac{d\theta(x)}{dx} \right)^2 - \alpha \frac{d\theta(x)}{dx} - \beta \cos{\theta(x)} \right], \label{energy_chm}
\end{equation}
where $\alpha = \tan^{-1}(|\mbox{\boldmath $D$}|/J),~\beta = 2\mu_B H_{\rm appl}/\mathcal{J}S,~$ and $\mathcal{J} = \sqrt{J^2 + D^2}$.
We minimize this energy in Eq. (\ref{energy_chm}) with respect to $\theta(x)$ and obtain the Sine-Gordon equation,
\begin{equation}
 \frac{d^2\theta(x)}{dx^2} - H^\ast \sin{\theta} = 0 \label{sine_gordon}
\end{equation}
$H^\ast = 2\mu_B H_{\rm appl}/(\xi_0^2S^2\sqrt{J^2+D^2})$ is a normalized applied magnetic field.
Then, we assume that the order of the lattice constant of the chiral helimagnet $a$ and that of the coherence length for the superconductor at zero temperature $\xi_0$ are same, $a = \xi_0$.
The solution of the equation (\ref{sine_gordon}) is 
\begin{equation}
 \theta(x) = 2\sin^{-1} \left[ {\rm sn} \left( \frac{\sqrt{H^\ast}}{k}x \right) \right] + \pi, \label{solution_theta}
\end{equation}
where $k$ is the modulus of Jacobi's elliptic function, ${\rm sn}(x|k)$.
$k$ is determined by the relation,
\begin{equation}
 \frac{\pi \alpha}{4\sqrt{H^\ast}} = \frac{E(k)}{k} \label{k_determination}
\end{equation}
Here, $E(k)$ is the complete elliptic integral of the second kind.
The relation between $k$ and $H^{\ast}$ is shown in Fig. \ref{k_appl}

\begin{figure}[t]
 \includegraphics[scale=0.3]{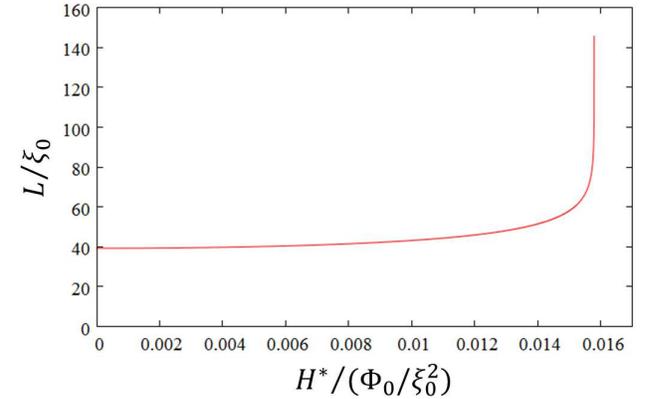}
 \caption{Dependence of the helical period on the applied magnetic field for $|\mbox{\boldmath $D$}|/J=0.16$.}
 \label{Fig_period}
\end{figure}

The helical period of the chiral helimagnet $L$ becomes large due to the applied magnetic field.
The relation between the helical period $L$ and the normalized applied magnetic field $H^\ast$ is given as,
\begin{equation}
 \frac{L}{\xi_0} = \frac{2kK(k)}{\sqrt{H^\ast}}. \label{period}
\end{equation}
$K(k)$ is the complete elliptic integral of the first kind and this relation (Eq. (\ref{period})) is shown in Fig. \ref{Fig_period}. 

The external magnetic field from the chiral helimagnet and the uniform applied magnetic field ($\mbox{\boldmath $H$}_{\rm ext})_z$ becomes as,
\begin{equation}
 (\mbox{\boldmath $H$}_{\rm ext})_z = H_0 \cos{\left[ 2\sin^{-1} \left\{ {\rm sn} \left( \frac{\sqrt{H^\ast}}{k}x~|~k \right) \right\} \right]} + H_{\rm appl}. \label{h_ext}
\end{equation}
Here, $H_0$ is a magnitude of the magnetic field from the chiral helimagnet.
This magnetic field $\mbox{\boldmath $H$}_{\rm ext}$ can be controlled by the applied magnetic field $H_{\rm appl}$.

\section{Result}
We show vortex states in superconductors under the magnetic field from the chiral helimagnet.

We solve Ginzburg-Landau equations (\ref{gl_fem_1})-(\ref{gl_fem_4}) and investigate stable vortex states.
We take the Ginzburg-Landau parameter $\kappa = \lambda_0 / \xi_0 = 10$, the temperature $T = 0.3T_c$.
The ratio of the Dzyaloshinsky-Moriya interaction to the ferromagnetic exchange interaction $|\mbox{\boldmath $D$}|/J$ is $0.16$, which is the experimental data of CrNb$_3$S$_6$ \cite{DM}.
The system size is $5.0L'\xi_0 \times 20\xi_0$, where $L'=L/\xi_0$ is a normalized helical period (see Eq. (\ref{period})).
The helical period increases due to the uniform applied magnetic field shown in Fig. \ref{Fig_period}.
When we calculate the Ginzburg-Landau equations, we set following boundary condition,
\begin{equation}
 \left| \left( \frac{\hbar}{i} \nabla + \frac{e}{c} \mbox{\boldmath $A$} \right) \psi \right| \cdot \mbox{\boldmath $n$} = 0,~~~~\mbox{\boldmath $A$} \cdot \mbox{\boldmath $n$} = 0. \label{boundary}
\end{equation}
In order to solve Eqs. (\ref{gl_fem_1})-(\ref{gl_fem_4}) numerically, we use randomly given order parameters as initial states.
Then, we calculate order parameters iteratively and obtain a convergent solution.

After we obtain distributions of order parameters from Eqs. (\ref{gl_fem_1})-(\ref{gl_fem_4}), we calculate the Ginzburg-Landau free energy in Eq. (\ref{gl_free}).
The Ginzburg-Landau free energy depends on the external magnetic field.
In order to investigate the dependences of the free energy on the magnetic field, we vary the magnitude of the magnetic field and solve Eqs. (\ref{gl_fem_1})-(\ref{gl_fem_4}) with the convergent solution as an initial state.
Stable vortex states has already discussed in the reference\cite{Fukui_free}, but we change the system size in order to confirm our previous results.
Moreover, we consider various vortex states and discuss their stability. 

First, we show vortex states under the magnetic field from the chiral helimagnet without uniform applied magnetic field $|H_{\rm appl}/(\Phi_0/\xi_0^2)| = 0.00$.
Then, we investigate how free energy depends on the number of vortices, the vortex positions and their structures.

\subsection{Stability of single vortex state: position dependence}
\begin{figure}[t]
\centering
 \includegraphics[scale=0.65]{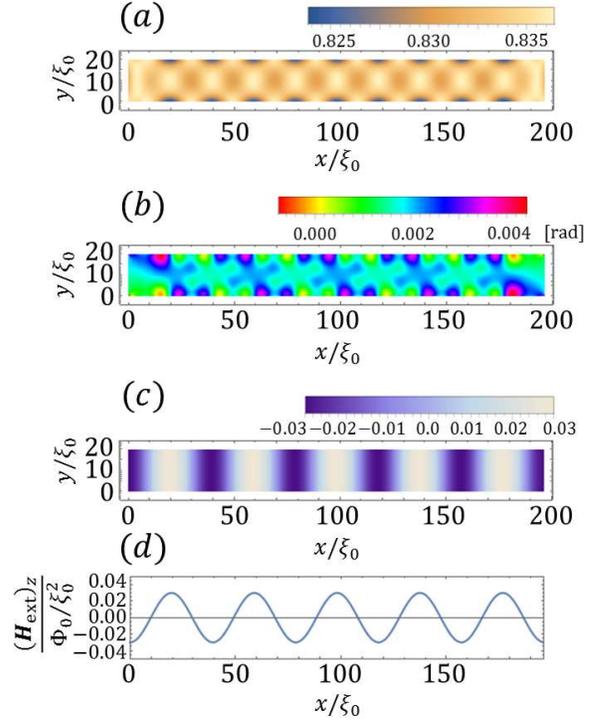}
 \caption{Distribution of (a) the amplitude of the order parameter normalized by the magnitude of the order parameter at zero field. (b) the phase $\varphi$ of the order paremter $\psi = |\psi|e^{i\varphi}$ and the magnetic field (c). (d) is the oscillating magnetic field at one $y$-coordinate.The amplitude of the oscillating magnetic field in Eq.(\ref{h_ext}) is $H_0/(\Phi_0/\xi_0^2) = 0.030$.}
 \label{vvortex_0}
\end{figure}

\begin{figure}[t]
\centering
 \includegraphics[scale=0.6]{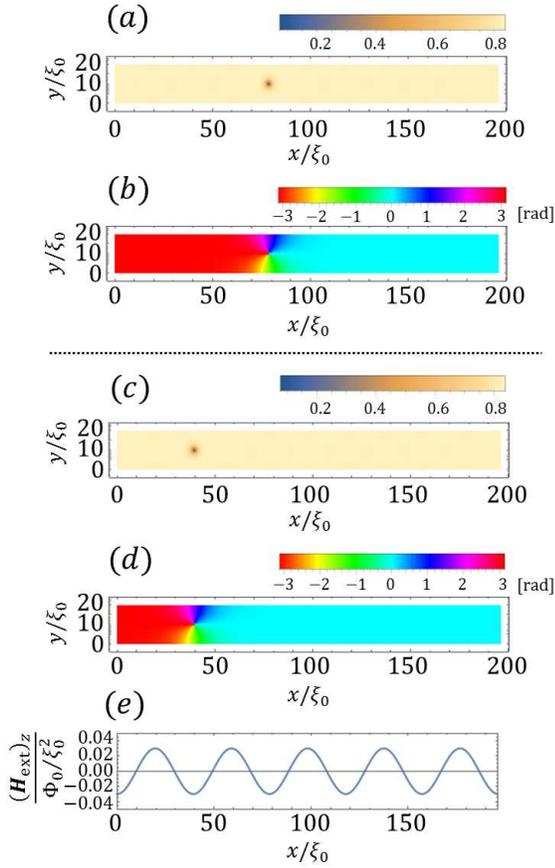}
 \caption{(a), (c) Distributions of the amplitude of the order parameter normalized by the magnitude of the order parameter at zero field. (b), (d) Distributions of the phase in the order parameter. (e) Distribution of the magnetic field. The amplitude of the oscillating magnetic field in Eq.(\ref{h_ext}) is $H_0/(\Phi_0/\xi_0^2) = 0.030$. A single vortex appears at $(x/\xi_0, y/\xi_0) \sim (80,10)$ [\ref{vvortex_1}(a) and \ref{vvortex_1}(b)], $(x/\xi_0, y/\xi_0) \sim $ $(40, 10)$ [\ref{vvortex_1}(c) and \ref{vvortex_1}(d)].}
 \label{vvortex_1}
\end{figure}

First, we investigate the stability of a no-vortex state (Fig. \ref{vvortex_0}) and single vortex states (Fig. \ref{vvortex_1}).
In Fig. \ref{vvortex_0}, we show distributions of the order parameter (Fig. \ref{vvortex_0}(a)), the phase of the order parameter (Fig. \ref{vvortex_0}(b)), the magnetic field (Figs. \ref{vvortex_0}(c) and \ref{vvortex_0}(d)).
We set the magnitude of the magnetic field in Eq. (\ref{h_ext}) as $H_0/(\Phi_0/\xi_0^2) = 0.030$.
In Fig. \ref{vvortex_0}(a), there is no vortex, but magnitudes of the order parameter at edges in the system oscillates due to the oscillating magnetic field. 
This oscillation occurs even if vortices appear.
On the other hand, Fig.\ref{vvortex_1} show distributions of the order parameters (Figs. \ref{vvortex_1}(a) and \ref{vvortex_1}(c)), phases (Figs. \ref{vvortex_1}(b) and \ref{vvortex_1}(d)), and the magnetic field (Fig. \ref{vvortex_1}(e)) for single vortex states.
Positions of the single vortex are $(x/\xi_0,~y/\xi_0) \sim (80, 10)$ in the Figs. \ref{vvortex_1}(a) and \ref{vvortex_1}(b), and $(x/\xi_0,~y/\xi_0) \sim (40, 10)$ in the Figs. \ref{vvortex_1}(c) and \ref{vvortex_1}(d).
These two single vortex states come from different initial states in iterative method.
In order to compare these vortex configurations of these states at several magnitudes of the field, we calculate the Ginzburg-Landau free energies (Eq. (\ref{gl_free})), which is shown in Table \ref{gl_one_compare}.
The state in Figs. \ref{vvortex_1}(a) and \ref{vvortex_1}(b) has a lower free energy than that of the states in Figs. \ref{vvortex_1}(c) and \ref{vvortex_2}(d).
But the energy difference is small $\Delta E/E \sim 10^{-4}$.
This difference may come from the boundary condition.
But we may say stability of single vortex state don't depend on the position of the vortex.


\begin{table}[htbp]
\begin{center}
\caption{Dependence of the magnetic field on two vortex states in Fig.\ref{vvortex_1}(a) and \ref{vvortex_1}(c)}
\begin{tabular}{|c|c|c|} \hline
$H_0/(\Phi_0/\xi_0^2)$ & $\mathcal{F}/\{(\alpha^2\xi_0^2)/\beta\}$ in Fig.\ref{vvortex_1}(a) & $\mathcal{F}/\{(\alpha^2\xi_0^2)/\beta\}$ in Fig.\ref{vvortex_1}(c) \\ \hline
 0.030                 & 48.86841  & 48.87938 \\ \hline
 0.035                 & 60.28517  & 60.29793 \\ \hline
 0.040                 & 73.67405  & 73.68857 \\ \hline
 0.045                 & 89.01415  & 89.03040 \\ \hline
 0.050                 & 106.28122 & 106.29918\\ \hline
\end{tabular}
\label{gl_one_compare}
\end{center}
\end{table}

\subsection{Stability of two-vortex states: interaction dependence}
\begin{figure}[t]
\centering
 \includegraphics[scale=0.6]{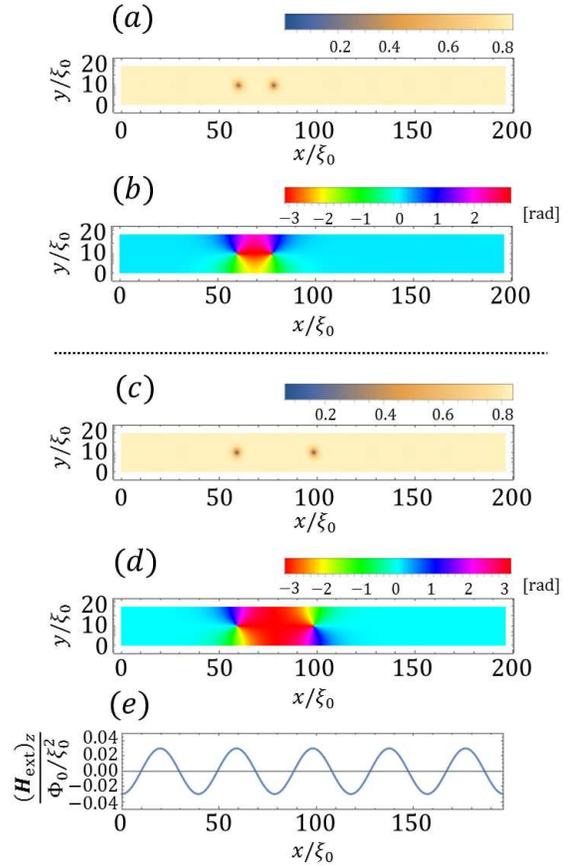}
 \caption{(a), (c) Distributions of the amplitude of the order parameter. (b), (d) Distributions of the phase in the order parameter. (e) Distribution of the magnetic field. The amplitude of the oscillating magnetic field in Eq.(\ref{h_ext}) is $H_0/(\Phi_0/\xi_0^2) = 0.030$. Two vortices appear at $(x/\xi_0, y/\xi_0) \sim (60,10)$ and $(80,10)$ [\ref{vvortex_2}(a) and \ref{vvortex_2}(b)], $(x/\xi_0, y/\xi_0) \sim $ $(60, 10)$ and $(100,10)$ [\ref{vvortex_2}(c) and \ref{vvortex_2}(d)].}
 \label{vvortex_2}
\end{figure}

Second, we investigate stability of two-vortex states.
When two or more vortices appear, vortices may interact with each other even in the oscillating field. 
We show two-vortex states in Fig. \ref{vvortex_2}.
Vortices are antiparallel (Figs. \ref{vvortex_2}(a) and \ref{vvortex_2}(b)) or parallel (Figs. \ref{vvortex_2}(c) and \ref{vvortex_2}(d)) with each other, respectively.
So, the interaction between vortices is different in these two states.
In Figs. \ref{vvortex_2}(a) and \ref{vvortex_2}(b), the interaction between vortices is attractive.
While in Figs. \ref{vvortex_2}(c) and \ref{vvortex_2}(d), the interaction between vortices is repulsive.
We compare Ginzburg-Landau energies of these states for several magnitudes of the field (Eq. (\ref{gl_free})), which is shown in Table \ref{gl_two_compare}.
From Table \ref{gl_two_compare}, we find the antiparallel vortex state is more stable than the parallel state for all field region.

\begin{figure}[t]
 \includegraphics[scale=0.45]{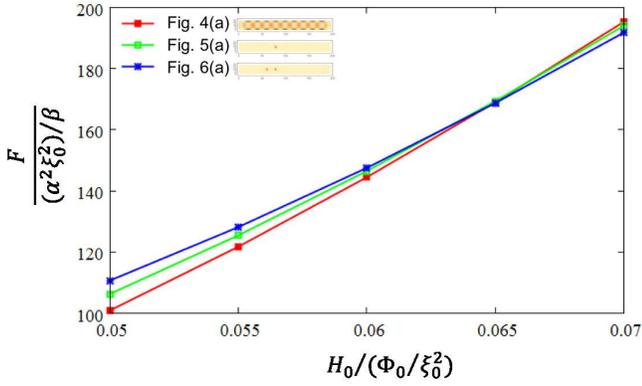}
 \caption{Dependences of free energies on the magnetic field in Fig.\ref{vvortex_0}, \ref{vvortex_1}, and \ref{vvortex_2}.}
 \label{ggl_free_0_1_2}
\end{figure}

Moreover, we compare the dependence of the free energies in no-vortex state (Fig. \ref{vvortex_0}(a)), single vortex state (Fig. \ref{vvortex_1}(a)) and two-vortex state (Fig. \ref{vvortex_2}(a)) on the magnetic field $H_0/(\Phi_0^2/\xi_0^2)$.
The dependence is shown in Fig. \ref{ggl_free_0_1_2}.
For smaller amplitude of the magnetic field, the no-vortex state is most stable.
Increasing the magnetic field, the single and the two-vortex states gradually becomes stable.
Further increasing the amplitude of the field, when the single-vortex state becomes more stable than the no-vortex state, two-vortex state becomes the most stable state.
So, vortices prefer to appear as a pair of up- and down-vortices.

\begin{table}[t]
\begin{center}
\caption{Dependence of the magnetic field on two vortex states in Fig.\ref{vvortex_2}}
\begin{tabular}{|c|c|c|} \hline
$H_0/(\Phi_0/\xi_0^2)$ & $\mathcal{F}/\{(\alpha^2\xi_0^2)/\beta\}$ in Fig.\ref{vvortex_2}(a) & $\mathcal{F}/\{(\alpha^2\xi_0^2)/\beta\}$ in Fig.\ref{vvortex_2}(c) \\ \hline
 0.030                 & 60.03972   & 61.15391 \\ \hline
 0.035                 & 69.76342   & 70.84697 \\ \hline
 0.040                 & 81.45461   & 82.51607 \\ \hline
 0.045                 & 95.09862   & 96.14271 \\ \hline
 0.050                 & 110.67555 & 111.70504\\ \hline
\end{tabular}
\label{gl_two_compare}
\end{center}
\end{table}

\subsection{Vortex number of the most stable state \label{number_stable}}
\begin{figure}[t]
\centering
 \includegraphics[scale=0.6]{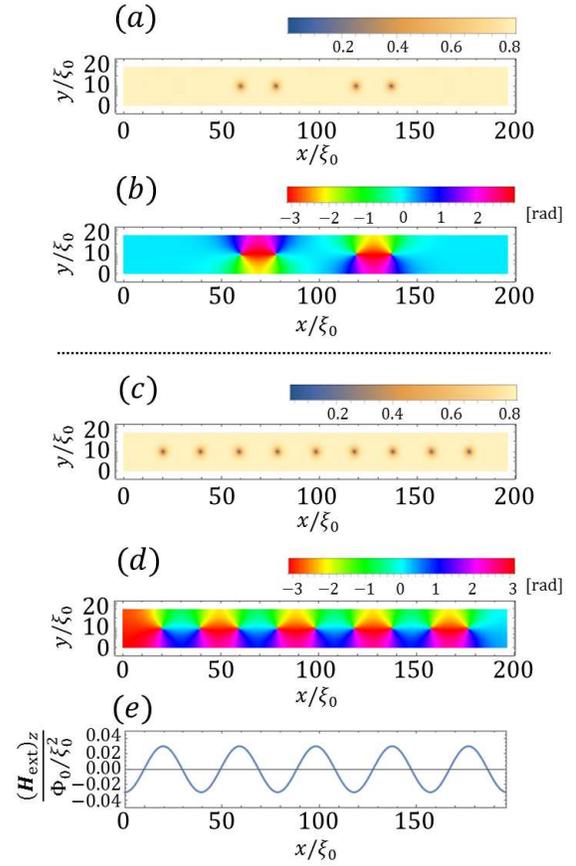}
 \caption{(a), (c) Distributions of the amplitude of the order parameter. (b), (d) Distributions of the phase in the order parameter. (e) Distribution of the magnetic field. The amplitude of the oscillating magnetic field in Eq.(\ref{h_ext}) is $H_0/(\Phi_0/\xi_0^2) = 0.030$. Four vortices appear at $(x/\xi_0, y/\xi_0) \sim (60,10), (80,10), (120, 10)$, and $(140,10)$ [\ref{vvortex_4_9}(a) and \ref{vvortex_4_9}(b)].
While nine vortices appear at every extremum point of the oscillating magnetic field, except for leftmost and rightmost points [\ref{vvortex_4_9}(c) and \ref{vvortex_4_9}(d)].}
 \label{vvortex_4_9}
\end{figure}

\begin{figure}[htbp]
 \includegraphics[scale=0.4]{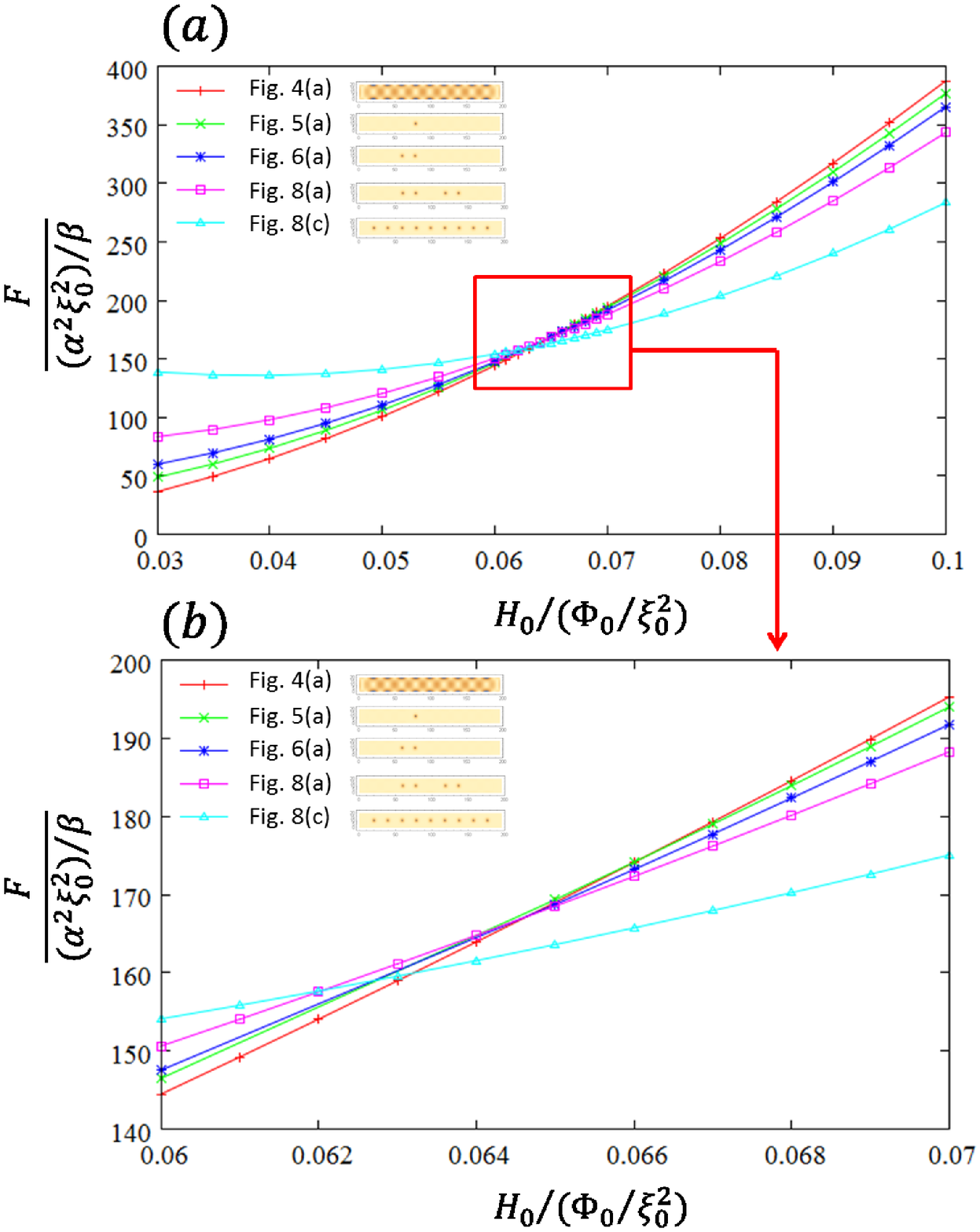}
 \caption{Dependences of free energies on the magnetic field in Figs. \ref{vvortex_0}, \ref{vvortex_1}, \ref{vvortex_2}, and \ref{vvortex_4_9}. The magnetic field region is (a) $0.030 \leq H_0/(\Phi_0/\xi_0^2) \leq 0.100$. (b) is a extended plot at (b) $0.060 \leq H_0/(\Phi_0/\xi_0^2) \leq 0.070$. }
 \label{ggl_free_4_9}
\end{figure} 

\begin{figure}[htbp]
 \includegraphics[scale=0.45]{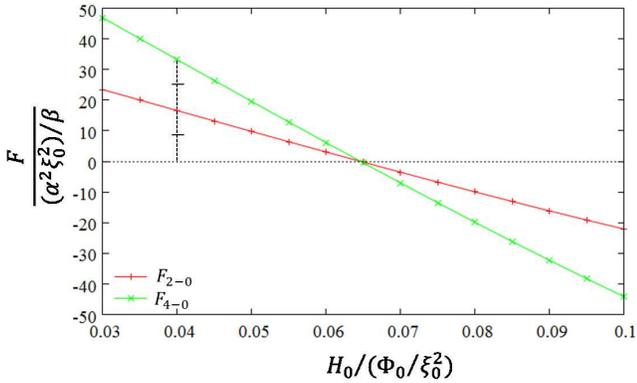}
 \caption{Differences of free energies in three vortex states in Figs. \ref{vvortex_0}, \ref{vvortex_2}(a), and \ref{vvortex_4_9}(a). $F_{2-0}$ represents the difference of free energies between zero vortex and one pair of vortices, and $F_{4-0}$ represents the difference of free energies between zero vortex and two pairs of vortices. }
 \label{ggl_free_diff_0_2_4}
\end{figure} 

\begin{figure}[htbp]
\centering
 \includegraphics[scale=0.6]{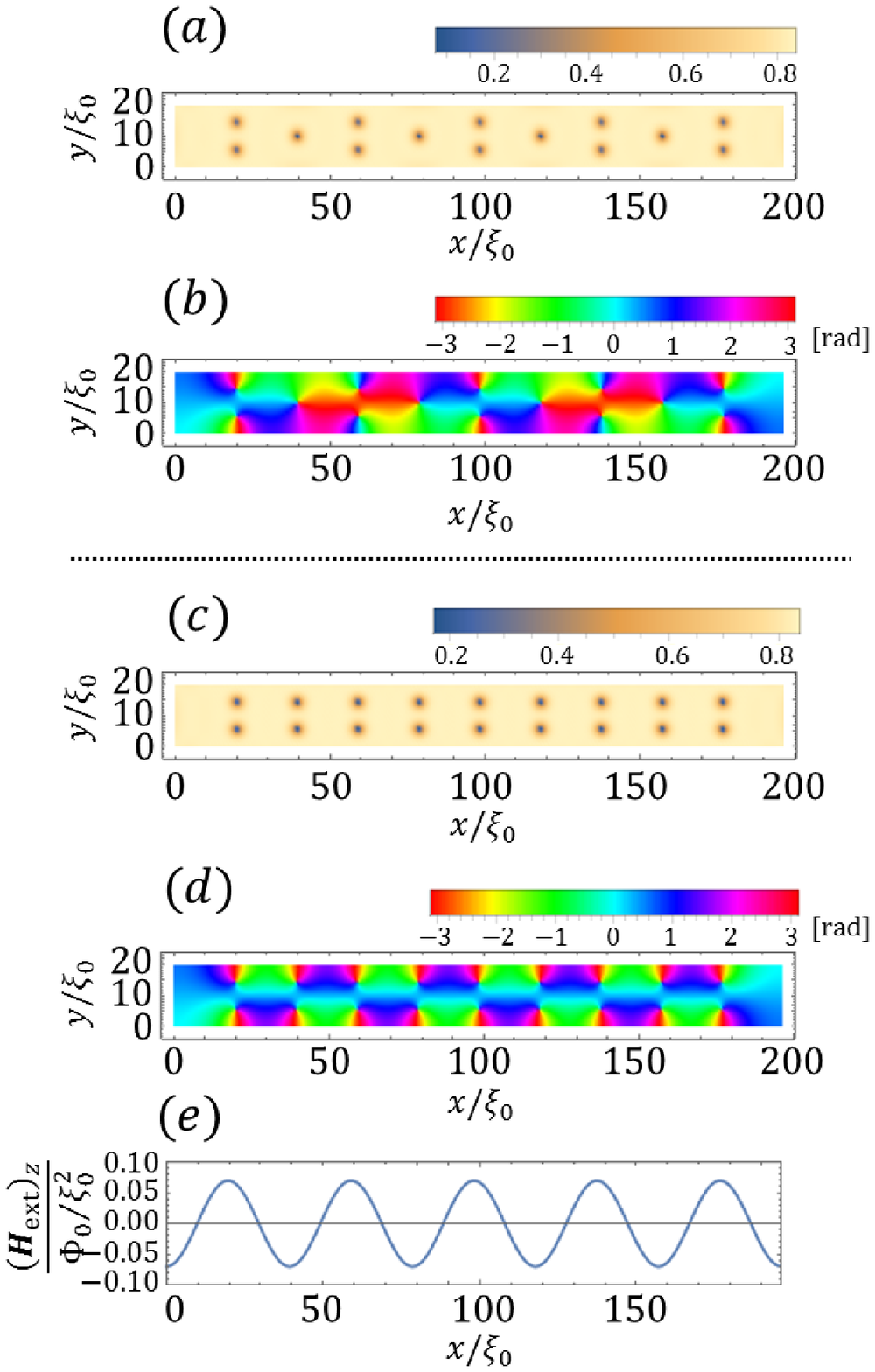}
 \caption{(a), (c) Distributions of the order parameter. (b), (d) Distributions of the phase in the order parameter. (e) Distribution of the magnetic field. The amplitude of the oscillating magnetic field in Eq.(\ref{h_ext}) is $H_0/(\Phi_0/\xi_0^2) = 0.030$. Two vortices appear in each magnetic fields $(\mbox{\boldmath $H$}_{\rm ext}/(\Phi_0/\xi_0^2))_z > 0$ and one vortex appears in each magnetic fields $(\mbox{\boldmath $H$}_{\rm ext}/(\Phi_0/\xi_0^2))_z < 0$ periodically [\ref{vvortex_2_1_2}(a) and \ref{vvortex_2_1_2}(b)].
While two vortices appear at all tops and bottoms of the oscillating magnetic field, except for edges [\ref{vvortex_2_1_2}(c) and \ref{vvortex_2_1_2}(d)].}
 \label{vvortex_2_1_2}
\end{figure}

\begin{figure}[htbp]
 \includegraphics[scale=0.4]{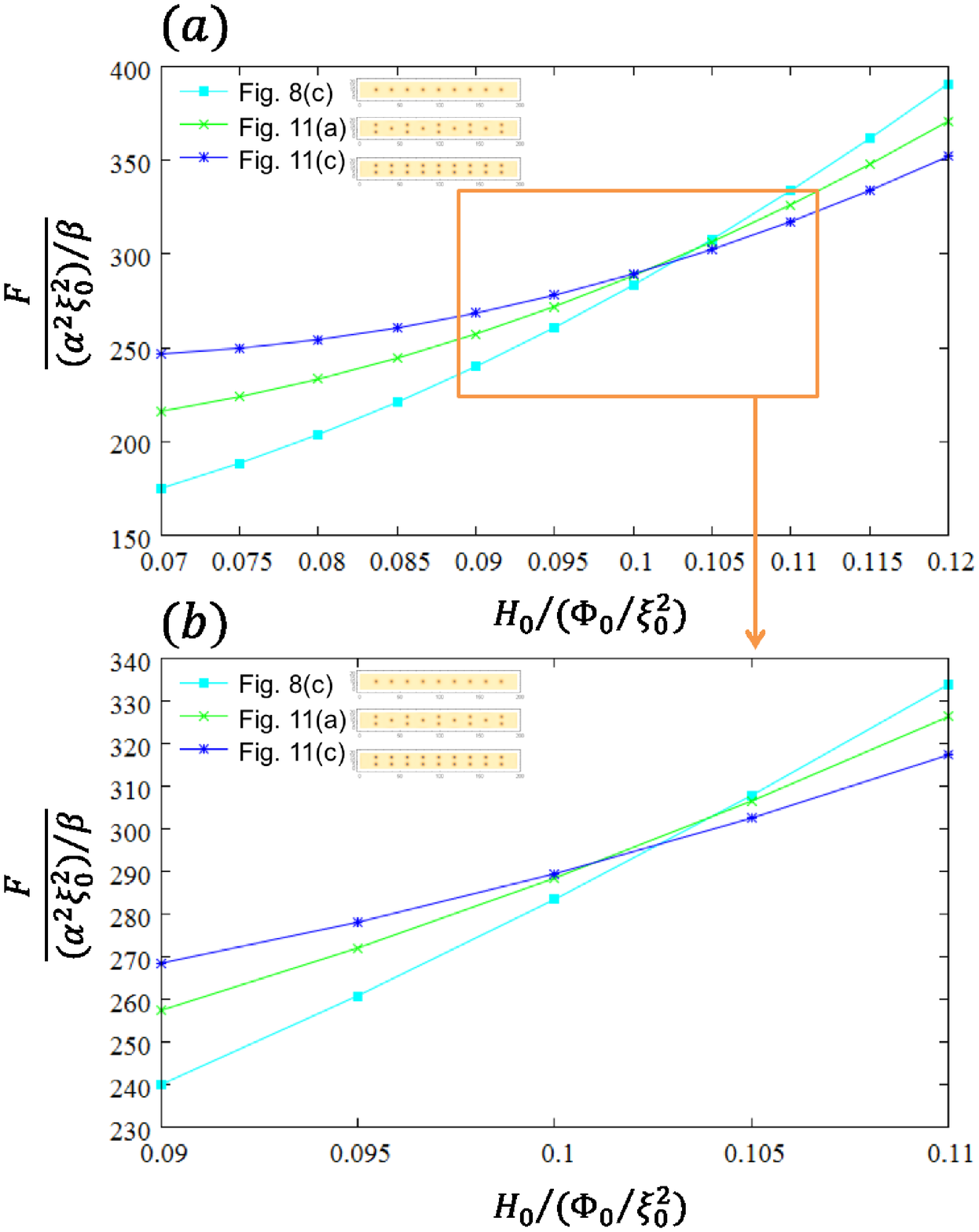}
 \caption{Dependences of free energies on the magnetic field in Fig.\ref{vvortex_4_9}(c), \ref{vvortex_2_1_2}(a), and \ref{vvortex_2_1_2}(c). The magnetic field region is (a) $0.070 \leq H_0/(\Phi_0/\xi_0^2) \leq 0.120$. (b) is a extended plot at (b) $0.090 \leq H_0/(\Phi_0/\xi_0^2) \leq 0.110$. }
 \label{ggl_free_over_9}
\end{figure} 

Third, we  show four- and nine-vortex states in Fig. \ref{vvortex_4_9}.
In Fig. \ref{vvortex_4_9}(a), two pairs of vortices appear.
On the other hand, in Fig. \ref{vvortex_4_9}(c), vortices appear at every extremum point of the oscillating magnetic field, except for leftmost and rightmost points.
In order to compare the stability, we also calculate free energies of vortex configurations in Figs. \ref{vvortex_4_9}(a) and \ref{vvortex_4_9}(c).
We show dependences of free energies on the magnetic field in Fig. \ref{ggl_free_4_9}.
In Fig. \ref{ggl_free_4_9}, we show free energies for the magnetic field region, (a) $0.030 \leq H_0/(\Phi_0/\xi_0^2) \leq 0.100$ and (b) $0.060 \leq H_0/(\Phi_0/\xi_0^2) \leq 0.070$.
We focus on free energies of zero-vortex (red), a pair of vortices state (blue), and two pairs of vortices state (purple).
Their differences of energies in three states for $0.03 \leq H_0 \leq 0.1$ are shown in Fig. \ref{ggl_free_diff_0_2_4}.
$F_{2-0}$ represents the difference of free energies between zero vortex and single-vortex-pair states, and $F_{4-0}$ represents the difference of free energies between zero vortex and two-vortex-pair states.
For this magnetic field range, $F_{4-0} \sim 2F_{2-0}$.
This means that $F_{2-0}$ is the free energy for a pair of vortices and $F_{4-0}$ is the free energies for independent two pairs of vortices.
There is a small difference $F_{4-0}-F_{2-0}$, which may come from the interaction between two pairs.

Finally, we consider the most stable vortex state among the no-, single-, two-, four-, and nine-vortex states.
Increasing the amplitude of the magnetic field $H_0/(\Phi_0/\xi_0^2)$, the state with the minimum free energy changes from no-vortex state (Fig. \ref{vvortex_0}(a)) to that nine-vortex state (Fig. \ref{vvortex_4_9}(c)).
So, the state with vortices in a row is the most stable state for higher field.
On the other hand, the single-vortex state [Fig. \ref{vvortex_1}(a)], the two-vortex state [Fig. \ref{vvortex_2}(a)], and four-vortex state [Fig. \ref{vvortex_4_9}(a)] do not become the minimum free energy state for whole field region.
Under the uniform magnetic field, the number of vortices increases one, two, three, and so on \cite{Kokubo_vortex}. 
In contrast to the uniform magnetic field case, vortices appear all at once in the oscillating magnetic field case.
In order to explain this result, we remember that for single-vortex states, the free energy do not depend on the position of the vortex (Table \ref{gl_one_compare}).
This means that a vortex is equally stable at the extremum points of the oscillating magnetic field.
Therefore, we can consider the whole superconductor as an ensemble of small superconductors, whose sizes are $1/2L'\xi_0\times20\xi_0$.
Because in each domain the same magnitude of the magnetic field is applied, same states are equally stable in all domains.
Therefore, vortices appear for the larger magnetic field, a single vortex appears in every small superconductor.
So, the most stable state changes from the no-vortex state to nine-vortex state.

We expect that similar situation is possible for further larger magnetic field case.
Then, we investigates stability of vortex structures shown in Fig. \ref{vvortex_2_1_2}.
The amplitude of the magnetic field $H_0/(\Phi_0/\xi_0^2) = 0.07$.
In Fig. \ref{vvortex_2_1_2}(a), two vortices appear in each maximum points and one vortex appears in each minimum points of the oscillation field.
On the other hand, in Fig. \ref{vvortex_2_1_2}(c), two vortices appear in all extremum points of the oscillation field.
From Fig. \ref{ggl_free_4_9}, the vortex structure in Fig.\ref{vvortex_4_9}(c) is most stable at $H_0/(\Phi_0/\xi_0^2) = 0.07$.
We compare free energies of vortex states in Figs. \ref{vvortex_4_9}(c), \ref{vvortex_2_1_2}(a), and \ref{vvortex_2_1_2}(c).
Their free energies are shown in Fig. \ref{ggl_free_over_9}.
From Fig. \ref{ggl_free_over_9}, we find that the structure with the minimum free energy changes from the structure in Fig. \ref{vvortex_4_9}(c) to that in Fig. \ref{vvortex_2_1_2}(c).
The structure in Fig. \ref{vvortex_2_1_2}(a) doesn't become the stable state in all magnetic field region.
In Fig. \ref{vvortex_2_1_2}(a), the number of the vortices in negative field regions is one, but the number of vortices in negative field region is two in Fig. \ref{vvortex_2_1_2}(c).
When the amplitude of the oscillating magnetic field is large and two vortices appear in a small superconducting region, in all of the regions there appear two vortices in all small regions all at once.

\subsection{Vortex states under a magnetic field composed of chiral and uniform applied magnetic fields}
\begin{figure}[t]
\centering
 \includegraphics[scale=0.6]{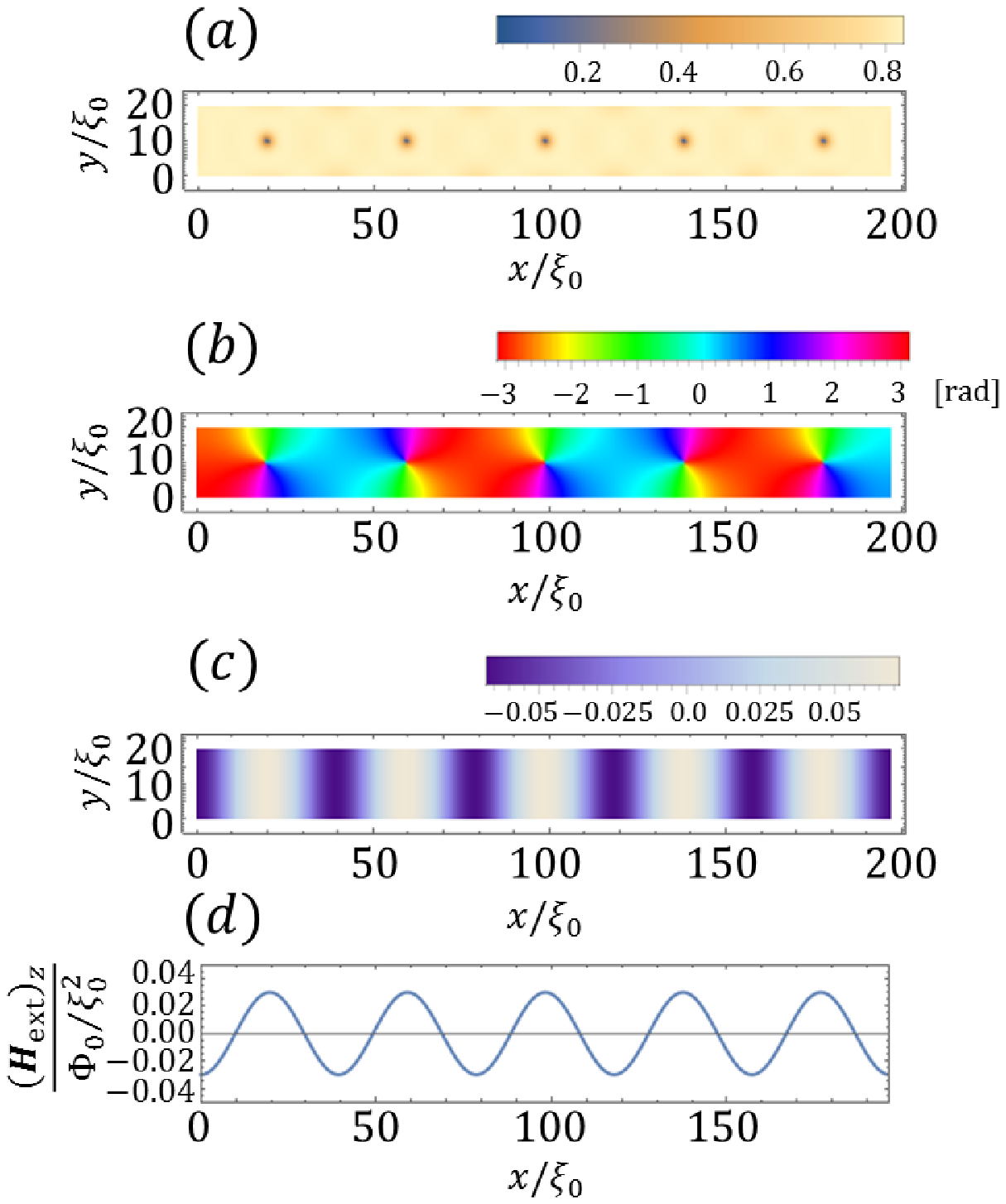}
 \caption{Distribution of (a) the order parameter, (b) the phase $\varphi$ of the order paremter $\psi = |\psi|e^{i\varphi}$ and the magnetic field (c). (d) is the oscillating magnetic field at one $y$-coordinate. Magnetic field from the chiral helimagnet is $H_0/(\Phi_0/\xi_0^2) = 0.070$ and the applied magnetic field is $H_{\rm appl}/(\Phi_0/\xi_0^2) = 0.0020$. Only vortices parallel to the magnetic field region $(H_{\rm ext}/(\Phi_0/\xi_0^2))_z > 0$ appear.}
 \label{vvortex_5_appl}
\end{figure} 
\begin{figure}[t]
 \includegraphics[scale=0.45]{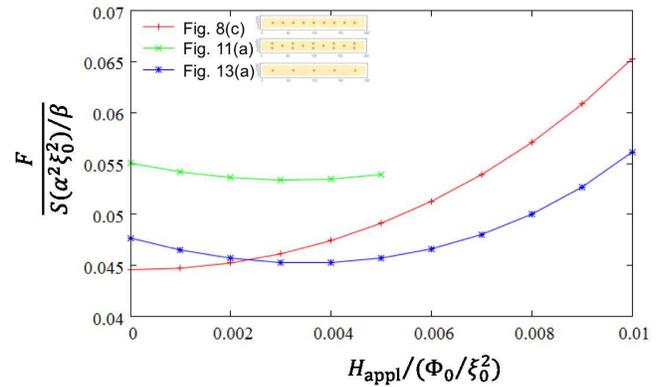}
 \caption{Dependences of free energies on the magnetic field in Fig.\ref{vvortex_4_9}(c), \ref{vvortex_2_1_2}(a), and \ref{vvortex_5_appl}. The magnetic field from the chiral helimagnet  is fixed for $H_0/(\Phi_0/\xi_0^2) = 0.070$. }
 \label{ggl_free_appl}
\end{figure}
In this sub-section, we show vortex states under a composite magnetic field with the chiral and uniform magnetic fields (Eq. (\ref{h_ext})).
When the uniform magnetic field is applied to the chiral helimagnet, the helical period of the chiral helimagnet changes as Eq. (\ref{period}).
For $H_0/(\Phi_0/\xi_0^2) = 0.070$ and $H_{\rm appl}/(\Phi_0/\xi_0) = 0.0020$ vortex structures are shown in Fig. \ref{vvortex_5_appl}.
In Fig. \ref{vvortex_5_appl}, there are no vortices antiparallel to the uniform magnetic field, which is called down-vortex.
We call the state in Fig. \ref{vvortex_5_appl} 1up-0down state.
Increasing the uniform magnetic field, the number of vortices parallel to the uniform magnetic field, which is called up-vortex, increases.
In lower uniform magnetic field, we think there is a possibility that vortex state in Fig. \ref{vvortex_2_1_2}(a) becomes stable.
We call the state in Fig. \ref{vvortex_2_1_2}(a) 2up-1down state.
As discussed in Subsection \ref{number_stable}, under only oscillating magnetic field increasing the amplitude of oscillating magnetic field $H_0/(\Phi_0/\xi_0^2)$, the most stable state changes from zero vortex state in Fig. \ref{vvortex_0} to nine-vortex state (1up-1down) in Fig. \ref{vvortex_4_9}(c), finally, to eighteen-vortex state (2up-2down) in two rows in Fig. \ref{vvortex_2_1_2}(c).
Then, in order to determine stabilities of these vortex states under the composite magnetic field, we compare the free-energies of vortex states.
We show free energies of 1up-1down, 2up-1down and 1up-0down states for fixed $H_0/(\Phi_0/\xi_0^2) = 0.070$, as functions of the uniform applied magnetic field in Fig. \ref{ggl_free_appl}.
We find that when the uniform magnetic field increases, the most stable state changes from the 1up-1down state to the 1up-0down state. 
The 2up-1down state has larger free energy in this field range. 
The critical field $H_{\rm appl}/(\Phi_0/\xi_0^2) \sim 0.02$ is much smaller than $H_0/(\Phi_0/\xi_0^2) = 0.07$. 
Therefore, we can say small uniform magnetic field easily remove down-vortices.

\section{Summary}
We have investigated vortex states in the two-dimensional superconductor under the helical magnetic field from the chiral helimagnet using two-dimensional Ginzburg-Landau equations.
We find that when the magnetic field from the chiral helimagnet increases, up and down-vortices appear all at once in all periodic regions. 
This behavior is different from that in a finite superconductor under the uniform magnetic field.
Under the composite magnetic field with the oscillating and uniform fields, increasing uniform magnetic field, from the 1up-1down vortex state down-vortices disappear.
This transition field is much smaller than the helical magnetic field amplitude $H_0/(\Phi_0/\xi_0^2)$.

In this study, we assume that the lattice constant $a$ and the coherence length $\xi_0$ are same.
The coherence length in a conventional superconductor may be longer than the lattice constant.
We believe that results are not much different when the lattice constant $a$ is smaller than $\xi_0$.
But in future, we will investigate how vortex states change when the coherence length and the lattice constant are different.
In addition, we consider the two-dimensional superconductor.
In the two-dimensional superconductor, we take only the $z$-component magnetic field from the chiral helimagnet $(\mbox{\boldmath $H$}_{\rm CHM})_z$.
In order to discuss effects of chirality in the chiral helimagnet completely, we take all components of helical magnetic field.
So, we should consider three-dimensional superconductor \cite{Fukui_3D}.
Such study is in progress.


\end{document}